\documentstyle[prl,tighten,aps]{revtex}
\def\beqa{\begin{eqnarray}}
\def\eeqa{\end{eqnarray}}
\def\beq{\begin{equation}}
\def\eeq{\end{equation}}

\def\bib#1{$^{\ref{#1}}$}

\def\eg{{\it e.g. }}

\def\a{\`a }% Accenti italiani
\begin{document}
\def\bib#1{[{\ref{#1}}]}
   \title{\normalsize \bf 
Phenomenological scaling laws relating the observed galactic \\ 
dimensions to Planck action constant}

\author{\normalsize 
Salvatore Capozziello$^{a,c,}$\thanks{E-mail:
capozziello@vaxsa.csied.unisa.it}, 
Salvatore De Martino$^{b,c,}$\thanks{E-mail:
demartino@physics.unisa.it},
Silvio De Siena$^{b,c,}$\thanks{E-mail:
desiena@physics.unisa.it}, and
Fabrizio Illuminati$^{b,c,}$\thanks{E-mail:
fabrizio@leopardi.phys.unisa.it} \\
{\small\em $^a$Dipartimento di Scienze Fisiche ``E. R. Caianiello",}\\ 
{\small\em $^b$Dipartimento di Fisica,}\\ 
{\small\em $^c$INFN, Sez. di Napoli and INFM, Unit\a di Salerno, }\\
{\small\em Universit\`a di Salerno, I-84081 Baronissi (SA), Italy.}} 
\date{February 24, 1999}
\maketitle

	      \begin{abstract}
It is shown that the characteristic observed radius,
velocity, and temperature of a typical galaxy
can be inferred from Planck action constant through
a phenomenological scaling law holding on all
cosmological scales.    
               \end{abstract}

\vspace{20. mm}

\vspace{20. mm}

One of the relevant open problems
of modern cosmology is that of
explaining the observed stability and sizes of galaxies. 
These properties should be deduced solely 
by assuming galaxies to be relaxed and virialized systems where 
gravity is the only overall interaction \bib{binney}. 

Such interaction is taken to be
Newtonian, while  
the confining potentials, due to the mutual
attractions of stars (and the other components as dust and gas clouds)
can have several forms depending on the mass distributions.
For example, logarithmic potentials well describe the
regular motion of stars before the onset of chaos \bib{saasfee}.

The problem is extremely involved since galaxies undergo 
environmental effects, being never 
completely isolated systems; they always belong to 
larger gravitationally bound systems like loose and tight groups, 
associations and clusters of galaxies, 
and the observational times are so short that the 
overall dynamics can only be extrapolated \bib{binney}, \bib{vorontsov}.

Futhermore, the dynamics of
galaxies must obviously be related to some 
cosmological model and, due to the
cosmological evolution of large scale structures, 
it is a widely shared feeling that one should observe
in present day dynamics some traces of the
primordial quantum perturbations \bib{kolb}, \bib{sakharov}.  
For this last reason, it seems meaningful to
ask for some quantum signature in the today observed 
galaxies \bib{kolb}, \bib{sakharov}.
The main goal, and at the same time the main difficulty,
is to provide a physical route connecting
the estremely large size of galaxies
($\sim 10$kpc) with the extremely small numbers 
of quantum mechanics
($h\sim 10^{-27}$ erg sec).

As a first step toward this aim we introduce in the present
work a simplified model of a galaxy composed of 
self--gravitating microscopic constituents (nucleons,
or, equivalently hydrogen atoms, possibly ionized)
affected by some suitable mechanical fluctuations. 
We perform a simple, semi--quantitative analysis
of such a model, exploiting as the only observational
input the number of nucleons contained in a typical galaxy.
The characteristic radius, velocities and temperature of
such model galaxies turn out to
be explicit functions of the Planck action constant
and of the number of microscopic constituents.

The surprising and intriguing fact is that the quantities
(characteristic radius, velocity and temperature) 
so deduced, numerically
coincide with those observed for a typical real galaxy, 
whose gravitational constituents are stars (not nucleons).

To understand the significance of this first finding we
analyse the observational data relative to
different large cosmological structures, and we 
obtain a remarkable scaling relation between the typical
units of action of the chosen granular gravitational components,
ranging from nucleons up to stars and galaxies.

In this way, the characteristic dimensions
of large scale cosmological structures appear, due to
this phenomenological 
scaling relation, independent of the scale of the
constituents considered. It is only needed 
that they depend on a minimal scale of action
which, in order of magnitude, turns out to
coincide with the Planck action constant.

These results point toward the existence of
a mechanism of macroscopic quantum coherence also for 
large scale gravitational
systems, in analogy with other well known collective
mesoscopic and macroscopic phenomena.

We start by considering the total action for a bound 
system (in our present case, a galaxy)
with a very large number $N$ of constituents.
Let $E$ be the total energy. 
Let ${\cal T}$ be the characteristic global 
time of the system 
(\eg the time in which a particle crosses 
the system, or the time in which
the system evolves and becomes relaxed). 
Combining these two quantities,
we get
\beq
\label{1}
A \cong E{\cal T} \, ,
\eeq

\noindent 
which is the total action. 
The only hypothesis which we need is that the 
system may undergo a time--statistical 
fluctuation, so that the 
characteristic time $\tau$ for the stochastic 
(chaotic) motion per particle will be
\bib{calogero}, \bib{demartino}
\beq
\label{2}
{\tau}\cong\frac{{\cal T}}{\sqrt{N}} \, .
\eeq

This hypothesis naturally emerges from the fact that 
if $N$ is large, the dynamics must be affected by some
kind of statistical (chaotic) fluctuation
\bib{binney}, \bib{contopoulos}.

We can then define an energy per granular component
\beq
\label{3}
\epsilon\cong\frac{E}{N} \, ,
\eeq

\noindent
so that the characteristic (minimal) unit of action 
$\alpha = \epsilon \tau$ per granular
component is expressed by the following crucial scaling
relation
\beq
\label{4}
\alpha=\epsilon\tau\cong\frac{A}{N^{3/2}} \, .
\eeq

Let us now consider the relevant data for galaxies, in order
of magnitude. 
The energy per unit of mass   
in a realistic galactic potential \bib{contopoulos}, is of the order 
$10^{15}$ (cm/sec$)^2$, the period of a galactic rotation,
which can be assumed as the characteristic global time, is 
of the order \bib{binney}
\beq
\label{5}
{\cal{T}}_{rot} \cong 10^{15}\mbox{sec} \, ,
\eeq

\noindent while the
total mass of a typical galaxy is of the order \bib{binney}
\beq
\label{6}
M \cong 10^{44}\mbox{gr} \, .
\eeq

From Eq.(\ref{1}), combining these numbers, we get
\beq
\label{7}
A \cong 10^{74}\mbox{erg sec} \, .
\eeq 

The number of nucleons in a galaxy is \bib{binney}
\beq
\label{8}
N \cong 10^{68} \, .
\eeq

Introducing these numbers inside Eq.(\ref{4}) we get, up to at most
an order of magnitude, that 
{\it the characteristic unit of action for a galaxy,
considered as an aggregate of nucleons, is of the order of
Planck action constant}, 
$h \sim 10^{-27}$ erg sec. We notice that also if 
dark matter is taken into account, the result does not change
dramatically since the mass to luminosity ratio is of the order
$10 \div 100$.

It is interesting to note that the values which we have used are on the 
boundary for the onset of chaos, as the galaxy is assumed stable.
In other words, the stability of the system appears to be related to the 
existence of a minimum, finite unit of action, and therefore,
ultimately, to quantum mechanics.

On these grounds, in the framework of our model
we move to infer a general, phenomenological
scaling law that connects microscopic scales to the global
dimensions of a typical galaxy, thus providing the rationale
to bridge the gap between the small numbers of quantum mechanics
and the large numbers of cosmology.

We start by noting that relation (4) together with the numerical
result $\alpha \cong h$ can be reformulated as a scaling relation
for the {\it mean} action per microscopic component $a \equiv A/N$:
\beq
\label{9}
a \cong h \sqrt{N} \, .
\eeq

We can then deduce that the fluctuative factor $\sqrt{N}$ supplies
the rescaling coefficient from the microscopic scales to the
characteristic macroscopic dimensions. The latter then appear to be
suitable mean field averages. For such a scaling to be
universal, it must hold in general, not only for actions.

In particular, it is crucial that it be verified also for lengths. 
Given the nucleons as the basic microscopic constituents in our model,
the natural quantum unit of length associated to each single constituent
is the Compton wavelength $\lambda_c = h/mc$, with $c$ the velocity
of light, and $m \cong m_{p} \cong 10^{-24}$ gr, the proton mass.    
Defining $R$, the macroscopic geometrical size of our model galaxy, we
must have
\beq
\label{10}
R \cong \lambda_{c} \sqrt{N} \, .
\eeq

\noindent 
With $N$ given by Eq.(\ref{8}), we obtain
\beq
\label{11}
R \cong 10^{21} \div 10^{22} \mbox{cm} \simeq 1 \div 10 \mbox{kpc} \, .
\eeq

\noindent 
But this is exactly the typical scale of length for a 
real galaxy. The numerical agreement of
formula (10) with the observed galactic radii, 
is interesting {\it per se}, independently of the present
derivation, since it
links the scale of a cosmological structure like a galaxy
to the Compton wavelength of
the elementary constituents (the nucleons) and to the
total number of such constituents. 

We further notice that Eq. (10) provides the correct order
of magnitude of the observed radii also if one considers
other large scale cosmological structures such as clusters
of galaxies or the whole Universe, provided one inserts the
correct value of the number of nucleons $N$ contained in such
structures. 
For example, in the case
of the Universe the current estimates provide $N \cong 10^{80}$
\bib{peebles}, and relation (10) yields then $R \cong 10^{28}$
cm, which coincides with the order of magnitude
of the observed radius of the Universe \bib{peebles}. 

This fact indicates that 
the quantum parameter $\lambda_{c}$ and the number of 
nucleonic constituents $N$
determine the real astrophysical dimensions.

It is interesting to notice that for typical galaxies $R$
is the characteristic dimension 
where their rotation curve can be assumed flat \bib{binney} and
where the halo and the disk stabilize each other. 

The scaling relations (4) and (10), and the characteristic global
scales of time of a typical galaxy also yield, as derived relations,
the order of magnitude of galactic velocities.
We can consider the period of rotation (5) or another significant
global time scale like the time ${\cal{T}}_{vir} \cong 10^{16}
\div 10^{17}$ sec after which the galaxy can be considered 
virialized \bib{binney}.

In each case, the expressions 
$v_{rot} = R/{\cal{T}}_{rot}$ and $v_{disp} = R/{\cal{T}}_{vir}$
cover the range of velocities of a typical galaxy, from the
dispersion velocity of stars ($\cong 10^{5}$cm/sec) to the circular
speed of a star in the disk ($\cong 10^{7}$cm/sec).

The phenomenological scaling relations (4) and (10) 
allow also to introduce thermodynamic estimates.
On the time scale associated to the virialization time 
${\cal{T}}_{vir}$, we can assume a well defined characteristic
temperature $T$ inside the galaxy. 
Having verified the numerical relation $\alpha \cong h$, we
can link the fluctuative microscopic time scale $\tau$ defined
in Eq. (2) to the temperature $T$ inside the galaxy by the
usual Planck--Boltzmann--Gibbs relation:
\beq
\label{12}
\tau\cong\frac{h}{k_{B}T} \, .
\eeq

\noindent
By Eq. (2), with the global time identified as the virialization
time ${\cal{T}}_{vir}$  we then obtain
\beq
\label{13}
T\cong \frac{h}{k_{B}} \frac{\sqrt{N}}{{\cal{T}}_{vir}} 
\, .
\eeq

Inserting the numerical values, Eq. (13) yields
\beq
\label{14}
T \cong 10^{6} \div 10^{7} \mbox{K} \, ,
\eeq

\noindent which is the characteristic range of temperatures
inside a star.

Summing up, by assuming a simplified model of a galaxy
composed by nucleons subject to a suitable mechanism of
chaoticity, we have derived values of the characteristic
galactic dimensions in agreement with the
observed values. This result is provided by a phenomenological
scaling relation, of apparent universal validity, which connects
microscopic and cosmological scales through a rescaling factor
depending only on the number $N$ of elementary, granular constituents.

One might wonder whether the above are all numerical coincidences
(which were not previously known), possibly artifacts of the
model we have considered, and not in 
fact expressions of a simple, but apparently general, 
scaling law.
A crucial objection seems to be that stars, rather than nucleons,
are the natural candidates as elementary
gravitational constituents of a typical galaxy.

This apparent difficulty can be solved exploiting a very
simple scaling law, which holds true on
any scale. Let us introduce the number $N_{s}$ of stars
contained in a typical galaxy, and the number $N_{ns}$
of nucleons in a star. We can then obviously write
for the total number of nucleons in a typical galaxy:
\beq
\label{15}
N \cong N_{s}N_{ns} \, .
\eeq 

\noindent Exploiting Eq. (15) and performing straightforward
manipulations we can write Eq. (10) in the following
form:
\beq
\label{16}
R \cong \lambda_{s} \sqrt{N_{s}} \, ,
\eeq

\noindent where 
\beqa
\label{17}
\lambda_{s} & \equiv & \frac{A_{s}}{M_{s}c} \, , \nonumber \\
&& \nonumber \\
A_{s} & \equiv & h N_{ns}^{3/2} \, , \\
&&  \nonumber \\
M_{s} & \equiv & m N_{ns} \, . \nonumber
\eeqa

\noindent
Here $M_{s}$ obviously coincides with
the total mass of a star, while the
quantity $A_{s}$ is the characteristic unit of action of 
a star in the framework of our model, taking the stars
as the elementary constituents of a typical galaxy.
Inserting the numerical values \bib{binney} $N_{ns} \cong 10^{57}$,
$N_{s} \cong 10^{11} \div 10^{12}$, we obtain
\beq
\label{18}
\lambda_{s} \cong 10^{14} \div 10^{15} \mbox{cm} \, ,
\eeq
 
\noindent
which is the typical range of interaction for a star
(e.g. that of the Solar System), while 
for $R$ we obviously obtain again the value (11).

Therefore, equations (10) and (16) show that we can derive 
the observed galactic radius $R$ either by considering a galaxy as a
gas of $N$ nucleons with the fluctuation (2) defined with respect
to $N$, or by considering, as usual, a typical
galaxy as a gas of stars and assuming the fluctuative
ansatz (2) rescaled respect to the number of stars $N_{s}$. 

The reason for the validity
of this relation of self--similarity (which in principle
holds on any scale) apparently relies on the existence of  
a minimal scale of action which is needed for mechanical
stability. In fact, the numerical value of 
the unit of action $A_{s}$ defined
in eq. (17) is $\cong 10^{58}$ erg s and thus coincides, in
order of magnitude, with the total action for a typical star
\bib{binney}, \bib{contopoulos}, \bib{peebles}.
Thus the rescaling relations (4) and (9) hold true also
for a star, and $\lambda_{s}$ appears as the effective
macroscopic Compton wavelength of a star.

The universality of the scaling law can be further tested.
Having verified it for the actions and for the lengths,
let us test it also for the velocities and for the temperatures.
We can rewrite the formula for the radius $R$ of the galaxy
in the following two equivalent forms:
\beq
\label{19}
R \cong \frac{h}{mv_{n}} \cong \frac{A_{s}}{M_{s}v_{s}} \, ,
\eeq

\noindent
where we have defined, respectively, the characteristic
velocity $v_{n} \cong c/\sqrt{N}$ of a nucleon in a galaxy,
and the characteristic velocity $v_{s} \cong c/\sqrt{N_{s}}$
of a star in a galaxy. Inserting the numerical value of
$N_{s}$ we have $v_{s} \cong 10^{4} \div 10^{5}$ cm/sec,
which agrees with the order of magnitude of the dispersion
velocities of stars.

Regarding temperatures we proceed easily as follows.
Inserting eq. (15) into eq. (13) we can reexpress the
characteristic temperature as:
\beq
\label{20}
T \cong \frac{a_{ns}}{k_{B}}\frac{\sqrt{N_{s}}}{{\cal{T}}_{vir}}
\, ,
\eeq

\noindent where $a_{ns} \equiv h\sqrt{N_{s}} \equiv
A_{s}/N_{ns}$ is the mean action per nucleon in a
star.

Summing up, we can draw the following conclusions:

1) Assuming the scaling ansatz (9), suggested by the
form (2) of the statistical fluctuation for a mechanical
system with a large number $N$ of constituents, and
by the numerical agreement (4) of the characteristic
unit of action per component $\alpha$ and the Planck action
constant $h$, we have derived estimates in order of magnitude
of the macroscopic geometric, kinematic and thermodynamic
quantities of a typical galaxy. In all cases our predictions
coincide with the experimentally observed quantities. 

We have assumed relation (9) as a {\it purely phenomenological
law} whose relevance is simply due to the remarkable
numerical agreement that it provides with respect to the observed
quantities. 
Since the only external data used as inputs are
the number of components $N$ and the fundamental constants
$h$ and $c$, such an agreement can be hardly considered 
an accidental coincidence. 

2) Moving from the above phenomenological setting to a more
theoretical framework, one should notice that position (2)
has been first put forward by F. Calogero \bib{calogero}
in the attempt to demonstrate a possible cosmic origin of
quantization ascribed to a ``universal tremor'' generated
by the complex, chaotic gravitational dynamics of the 
nucleons in the Universe. 

However, we have verified that positions (2), (4), (9) 
hold also for galaxies, and not only for
the whole Universe. Further, we have shown that the scaling
relations (4) and (9) are valid also for stars, whose
mechanical equilibrium cannot certainly be attributed only
to gravitational interactions. In fact, we have shown
\bib{demartino}, \bib{noi} that the analysis introduced
in the present paper for cosmological structures also
applies to a wide range of stable, confined macroscopic
systems not only of gravitational nature.

Therefore we believe that the meaning of our numerical
findings is that the existence of the Planck action constant
$h$ and the velocity of light $c$, that is quantum field
theory, determine the stability and the 
characteristic dimensions of all macroscopic
systems, including large scale cosmological structures like
for instance galaxies. 
This happens through a simple
scaling law and a mechanism of self--similarity. 
What is most intriguing is that the
emergence of this quantum
signature seems not only to be a weak footprint of primordial
quantum fluctuations, but rather a strong conditioning of
macroscopic quantum cohrence on the today observed large
scale cosmological structures.

3) Obviously it would be desirable to derive the scaling 
law (9) from first principles, perhaps from suitable
mean field averages or semiclassical approximations on
the many--body quantum dynamics.

\vspace{2. cm}

\begin{centerline}
{\bf REFERENCES}
\end{centerline}
\begin{enumerate}
\item\label{binney}
J. Binney and S. Tremaine, {\it Galactic Dynamics}
(Princeton University Press, Princeton, 1987).
\item\label{saasfee}
J. Binney, J. Kormendy, S.D.M. White, 
{\it Morphology and Dynamics of Galaxies}\\
12th Adv. Course of Swiss Society of Astr. and Astroph. Saas-Fee 1982\\
Eds. L. Martinet and M. Mayor, Geneva Observatory Editions.
\item\label{vorontsov}
B.A. Vorontsov--Vel'yaminov, {\it Extragalactic Astronomy}
Harwood Academic Pub. (London) 1987. 
\item\label{kolb}
E. W. Kolb and M. S. Turner, {\it The Early Universe}
(Addison--Wesley, New York, 1990).
\item\label{sakharov}
A. Sakharov, Zh. Eksp. Teor. Fiz. {\bf 49}, 245 (1965).\\
V. F. Mukhanov, H. A. Feldman and R. H. Brandenberger, 
Phys. Rep. {\bf 215}, 203 (1992).
\item\label{calogero}
F. Calogero, Phys. Lett. {\bf A 228}, 335 (1997).
\item\label{demartino}
S. De Martino, S. De Siena and F. Illuminati,
Mod. Phys. Lett. {\bf B 12}, 291 (1998).
\item\label{contopoulos}
G. Contopoulos, N. K. Spyrou, and L. Vlahos (Eds.),
{\it Galactic Dynamics and N--Body Simulations},
Lecture Notes in Physics {\bf 433} (Springer Verlag,
Berlin--Heidelberg, 1994).
\item\label{peebles}
P. J. Peebles, {\it Principles of Physical Cosmology}
(Princeton University Press, Princeton, N. J., 1993).
\item\label{noi}
S. De Martino, S. De Siena and F. Illuminati, {\it Inference 
of Planck action constant by a classical fluctuative
postulate holding for stable microscopic and macroscopic
dynamical systems}, LANL e--print quant-ph/9901045 (1999). \\
S. Capozziello, S. De Martino, S. De Siena and F. Illuminati,
{\it Quantum Signature of Large Scale Cosmological Structures},
LANL e--print gr--qc/9809053 (1998). 
\end{enumerate}
 
\end{document}